\documentclass[fleqn,usenatbib]{mnras} 

\usepackage{amsmath}	
\usepackage{amssymb}	
\usepackage{tikz}

\usepackage{lipsum}                  
\usepackage[normalem]{ulem}          

\usepackage{graphicx}
\usepackage{caption}
\usepackage{subcaption}

\usepackage{tikz,xcolor,hyperref}

\definecolor{lime}{HTML}{A6CE39}
\DeclareRobustCommand{\orcidicon}{%
	\begin{tikzpicture}
	\draw[lime, fill=lime] (0,0) 
	circle [radius=0.16] 
	node[white] {{\fontfamily{qag}\selectfont \tiny ID}};
	\draw[white, fill=white] (-0.0625,0.095) 
	circle [radius=0.007];
	\end{tikzpicture}
	\hspace{-2mm}
}

\foreach \x in {A, ..., Z}{%
	\expandafter\xdef\csname orcid\x\endcsname{\noexpand\href{https://orcid.org/\csname orcidauthor\x\endcsname}{\noexpand\orcidicon}}
}

\title[mHz QPOs detection in 4U~1730--22 with NICER]{Detection of millihertz quasi-periodic oscillations in the low-mass X-ray binary 4U~1730--22 with NICER}

\author[G. C. Mancuso et al.]{G. C. Mancuso{\orcidA{}},$^{1,2}$\thanks{E-mail: gmancuso@iar.unlp.edu.ar} D. Altamirano{\orcidB{}},$^{3}$ P.~Bult{\orcidC{}},$^{4,5}$ J. Chenevez{\orcidD{}},$^{6}$
\newauthor
S. Guillot{\orcidE{}},$^{7}$ T. G\"uver{\orcidF{}},$^{8,9}$ G.~K.~Jaisawal{\orcidG{}},$^{6}$ C.~Malacaria{\orcidH{}},$^{10}$ M. Ng{\orcidI{}},$^{11}$
\newauthor
A.~Sanna{\orcidJ{}},$^{12}$ and T. E. Strohmayer{\orcidK{}}$^{13}$ \\
$^{1}$Instituto Argentino de Radioastronom\'{\i}a (CCT-La Plata, CONICET; CICPBA), C.C. No. 5, 1894 Villa Elisa, Argentina \\
$^{2}$Facultad de Ciencias Astron\'omicas y Geof\'{\i}sicas, Universidad Nacional de La Plata, Paseo del Bosque s/n, 1900 La Plata, Argentina \\
$^{3}$Physics \& Astronomy, University of Southampton, Southampton, Hampshire SO17 1BJ, UK \\
$^{4}$Department of Astronomy, University of Maryland, College Park, MD 20742, USA \\
$^{5}$Astrophysics Science Division, NASA Goddard Space Flight Center, Greenbelt, MD 20771, USA \\
$^{6}$DTU Space, Technical University of Denmark, Elektrovej 327-328, DK-2800 Lyngby, Denmark \\
$^{7}$Institut de Recherche en Astrophysique et Plan\'{e}tologie, UPS-OMP, CNRS, CNES, 9 avenue du Colonel Roche, BP 44346, F-31028\\ Toulouse Cedex 4, France \\
$^{8}$Istanbul University, Science Faculty, Department of Astronomy and Space Sciences, Beyaz\i t, 34119, \.Istanbul, Turkey \\
$^{9}$Istanbul University Observatory Research and Application Center, Istanbul University 34119, \.Istanbul Turkey \\
$^{10}$International Space Science Institute (ISSI), Hallerstrasse 6, 3012 Bern, Switzerland \\
$^{11}$MIT Kavli Institute for Astrophysics and Space Research, Massachusetts Institute of Technology, Cambridge, MA 02139, USA \\
$^{12}$Dipartimento di Fisica, Universit\`a degli Studi di Cagliari, SP Monserrato-Sestu km 0.7, 09042 Monserrato, Italy \\
$^{13}$Astrophysics Science Division and Joint Space-Science Institute, NASA's Goddard Space Flight Center, Greenbelt, MD 20771, USA \\
}

\begin{document}
%
%
%
\maketitle
\label{firstpage}
\begin{abstract}

We report the discovery of millihertz quasi-periodic oscillations (mHz QPOs) from the neutron star (NS) low-mass X-ray binary 4U~1730--22 using the Neutron Star Interior Composition Explorer (NICER). After being inactive for almost 50 years, 4U~1730--22 went into outburst twice between June and August 2021, and between February and July 2022. We analyse all the NICER observations of this source, and detect mHz QPOs with a significance $>$\,4\,$\sigma$ in 35 observations. The QPO frequency of the full data set ranged between $\sim$\,4.5 and $\sim$\,8.1 mHz with an average fractional rms amplitude of the order of $\sim$\,2\%. The X-ray colour analysis strongly suggests that 4U~1730--22 was in a soft spectral state during the QPO detections. Our findings are consistent with those reported for other sources where the mHz QPOs have been interpreted as the result of a special mode of He burning on the NS surface called marginally stable nuclear burning (MSNB). We conclude that the mHz QPOs reported in this work are also associated with the MSNB, making 4U~1730--22 the eighth source that shows this phenomenology. We discuss our findings in the context of the heat flux from the NS crust.

\end{abstract}

\begin{keywords}
accretion, accretion discs $-$ stars: individual: 4U 1730$-$22 $-$ stars: neutron $-$ X-rays: binaries.
\end{keywords}

%
%
%
%

\section{Introduction}\label{sec:intro}

A low-mass X-ray binary (LMXB) consists of a compact object, either a neutron star (NS) or a black hole (BH), accreting matter from a low-mass donor star (usually with a mass $\lesssim$~1\,$M_{\odot}$) through an accretion disc (see, e.g., \citealt{pringle1972}; \citealt{shakura1973}). LMXBs are mainly classified into persistent and transient X-ray sources. Persistent systems are continuously active, generally showing luminosities in the range 10$^{36-38}$ erg s$^{-1}$. On the other hand, transient sources exhibit periods of high X-ray luminosities or outbursts ($L_{\rm X} \simeq 10^{34-39}$ erg s$^{-1}$), which typically last for several weeks up to months (e.g., \citealt{tetarenko2016}; \citealt{bahramian2022} for a review), followed by long intervals (generally months to decades) of low luminosity ($L_{\rm X} \lesssim 10^{34}$ erg s$^{-1}$; see, e.g., \citealt{chen1997}; \citealt{lasota2001}; \citealt{Hameury2020} for reviews), with little or no mass accretion. 

One of the features that distinguishes NS from BH LMXBs is the presence of thermonuclear (type-I) X-ray bursts. Type-I X-ray bursts manifest as an initial fast ($\lesssim$\,10 sec) increase of the X-ray flux due to an explosive fusion of the accreted H and/or He into heavier elements (see, e.g., \citealt{lewin1993}; \citealt{strohmayer2006}; \citealt{galloway-keek2021} for reviews). After reaching the peak flux, the decay phase follows a power law (\citealt{intZand2014}) due to the cooling of the NS atmosphere; the decay can last between tens to a few hundreds of seconds.
 
Most NS LMXBs display aperiodic X-ray variability in their light curves during outbursts at different timescales from seconds to milliseconds. This variability is categorised as quasi-periodic oscillations (QPOs) or broad-band noise, based on its characteristics (see, e.g., \citealt{vanderklis2006} for a review). QPOs stand out as relatively sharp and narrow (with a quality factor Q $>$ 2) peaks in Fourier power density spectra, and are commonly well represented by a Lorentzian profile (e.g., \citealt{vanderKlis1989}; \citealt{vanderklis2000}; \citealt{belloni2002}). In this paper, we focus on QPOs with frequencies of the order of millihertz (mHz). Reported for the first time in three systems (4U~1636--53, 4U~1608--52, and Aql~X--1) by \citet{revnivtsev2001}, and later on, in other sources, viz. IGR~J17480--2446 (\citealt{linares2010}), 4U~1323--619 (\citealt{strohmayer2011}), IGR~J00291+5934 (\citealt{ferrigno2017}), GS~1826--238 (\citealt{strohmayer2018}), EXO~0748--676 (\citealt{Mancuso2019}), and 1RXS~J180408.9--342058 (\citealt{tse2021}), these QPOs are different from those generally found at higher frequencies. The properties of the mHz QPOs observed in seven of the aforementioned systems (4U~1636--53, 4U~1608--52, Aql~X--1, 4U~1323--619, GS~1826--238, EXO~0748--676, and 1RXS~J180408.9--342058; from now on, group 1) are similar (although with some exceptions, in particular, for the case of 1RXS~J180408.9--342058). For instance: (i) the mHz QPOs are only detected in a specific range of X-ray luminosities, $L_{2-20\, \rm keV} \simeq (0.05-3.5) \times 10^{37}$~erg~s$^{-1}$ (see Table 1 in \citealt{tse2021});\footnote{It is important to note that the lower limit of the luminosity at which mHz QPOs have been detected must be taken with caution, given that it was derived using a poorly constrained estimated distance. If we exclude this value, the lower limit for the mHz QPOs detectability would be $\sim 0.5 \times 10^{37}$ erg s$^{-1}$.} (ii)~they are intimately related to type-I X-ray bursts, vanishing after the occurrence of a burst; (iii) their fractional rms amplitudes are of the order of $\sim$\,1--4\,\% and decrease from $\sim$ 2--3\,\% at $\sim$\,2.5~keV down to $\sim$~0.2--1\,\% at $\sim$\,5 keV (e.g., \citealt{revnivtsev2001}; \citealt{altamirano2008mQPOs}; \citealt{lyu2020}). In addition, in two sources of group 1, an increase of the rms amplitude from $\sim$\,0.2 keV to $\sim$\,2.5 keV was reported.
Finally, a systematic frequency drift of the mHz QPOs was also observed in at least three sources in group 1 (\citealt{altamirano2008mQPOs}; \citealt{Mancuso2019}; \citealt{mancuso2021}). 

\citet{revnivtsev2001} associated the mHz oscillations with a special mode of nuclear burning on the NS surface. The findings of \citet{Yu2002} supported this suggestion, who found that the frequency of the kHz QPO and the 2--5\,keV flux variation due to the mHz QPO were anti-correlated in 4U~1608--52. This anti-correlation is compatible with an enlargement of the inner disc radius driven by the flux excess from the NS surface during each mHz QPO cycle. 

The features of the other two sources which showed mHz oscillations differ from those observed in group 1. \citet{linares2010,linares2012} found QPOs with frequencies of $\sim$\,2.8--4.5\,mHz in the 11 Hz pulsar IGR~J17480--2446. The luminosity at which the oscillations were seen was an order of magnitude higher ($L_{2-50 \rm keV}$\,$\sim$\,$10^{38}$\,erg s$^{-1}$) than in group 1. Furthermore, the authors reported a smooth light curve evolution from thermonuclear bursts to mHz QPOs at the outburst peak as the persistent flux increased and vice versa. This is at odds with the mHz QPOs observed in group 1, in which the oscillations and the bursts can appear at the same flux level. \citet{ferrigno2017} found an $\sim$\,8 mHz QPO in the accreting millisecond X-ray pulsar IGR~J00291+5934. These oscillations, however, occurred at a luminosity of $\sim 10^{36}$~erg~s$^{-1}$, i.e., an order of magnitude lower compared with the QPOs observed in group 1, and were undetectable at energies $\gtrsim$\,2\,keV; the nature of these QPOs is still very uncertain.

Using numerical simulations, \citet{heger2007} were able to reproduce the main characteristics of the mHz QPOs. In effect, their model predicts that, at the transition between the stable and non-stable regimes, a special oscillatory mode of burning is expected, with a period of $\approx$ 100 s. This burning regime was found to happen in a narrow range of mass accretion rates, explaining why the mHz QPOs are only detected at a specific X-ray luminosity. The physical mechanism responsible for the oscillations proposed by \citet{heger2007} is the marginally stable nuclear burning (MSNB) of He on the surface of an NS. Some of the features derived from the model, however, are inconsistent with observations. For instance, while the simulations indicated that the oscillations should appear near the Eddington mass accretion rate, $\dot{M}_{\rm Edd}$, observations put this value close to $\sim 10\%$ $\dot{M}_{\rm Edd}$ (e.g., \citealt{revnivtsev2001}). \citet{heger2007} argued that this discrepancy would disappear if the accreted gas were confined to $\sim$\,10\% of the NS surface as the accretion rate per unit area would be close to the Eddington rate. Yet the possible process behind this confinement of the material is still unclear. The suggestion of \citet{heger2007} is supported by the findings of \citet{lyu2016}, who found that the type-I X-ray bursts that followed the mHz QPOs in 4U~1636--53 had positive convexities in their light curves, a characteristic associated with the bursts taking place at the NS equator (\citealt{cooper2007}; \citealt{maurer2008}). Therefore, this could be evidence that the local mass accretion rate is higher at the equator. \citet{keek2009} proposed a different explanation for the discrepancy between the predicted and observed mass accretion rate required for triggering the mHz QPOs. They noticed that combining a turbulent rotational mixing of the accreted material and a higher heat flux from the crust might explain the observed accretion rate value at which the mHz QPOs are found.

%
%
%
%
\subsection{4U~1730--22}

4U~1730--22 is a transient LMXB detected for the first time in 1972 with the Uhuru observatory (\citealt{cominsky1978}). The system was in outburst for approximately 200 days, until it turned into quiescence (\citealt{cominsky1978}; \citealt{chen1997}). Since then, 4U~1730--22 has not been detected in outburst again until June 2021 (after nearly 50 years). At the beginning of the outburst, it was identified by MAXI/GSC as a potentially new X-ray transient, with the tentative name MAXI~J1733--222 (\citealt{kobayashi2021Atel}). However, it was soon noted after follow-up observations with the Neil Gehrels Swift Observatory (\citealt{gehrels2004}) that the outburst came from a position 0.8 arc-seconds away from CXOU J173357.5--220156, the source identified by \citet{Tomsick2007} as the quiescent counterpart of 4U~1730--22 (\citealt{kennea2021ATel-2}). This demonstrated that 4U~1730--22 was the origin of the outburst. After a rapid brightening of the source in July 2021 (\citealt{iwakiri2021ATel}), the Neutron Star Interior Composition Explorer (NICER) observed for the first time a thermonuclear X-ray burst, unambiguously establishing the nature of the compact object as an NS (\citealt{bult2021ATel}), as was previously suggested by \citet{Tomsick2007}. 
Using photospheric radius expansion bursts observed in the 2022 outburst, \citet{Bult2022} obtained a source distance estimate of 6.9 $\pm$ 0.2 kpc. Burst oscillations at a frequency of 584.65 Hz were also found in one of the bursts detected in the 2022 outburst (\citealt{Li2022}). 

In this paper, we report the discovery of mHz QPOs in the NS LMXB 4U~1730--22 during its last 2022 outburst, and discuss its likely association in the context of the marginally stable nuclear burning theory.

%
%
%
%

\vspace{-0.25cm}
\section{Observations and data analysis}\label{sec:dataanalysis}

Launched in June 2017 and installed on the International Space Station, NICER (\citealt{gend-arzo2017}) is a soft X-ray telescope. NICER X-ray Timing Instrument (XTI; \citealt{gendreau2016}) consists of 56 coaligned X-ray concentrator optics, each paired with a silicon-drift detector (\citealt {prigozhin2012}). With 52 out of the 56 detectors operating, NICER covers the 0.2--12 keV energy range, and provides an effective area of $\approx$ 1900 cm$^2$ peaking at 1.5 keV.

NICER observed 4U~1730--22 during its last outburst starting on 14 February 2022. In this paper, we analysed data until 18 August 2022, distributed among 136 observations. Each observation consists of 1 to 10 data segments. We also analysed for completeness the observations of the previous outburst of the source (under the obsIDs 42022001mm, with mm ranging from 01 to 34), i.e., the one that took place between early June and late August 2021.

We processed the NICER observations using HEASOFT version 6.30.1 and NICERDAS version 9, along with the calibration database (CALDB) version 20210707. We filtered the data applying the standard screening criteria, i.e., pointing offset $<$ 54 arcsec, bright Earth limb angle $>$~30$^{\circ}$, dark Earth limb angle $>$ 15$^{\circ}$, and outside the South Atlantic Anomaly, via the task \texttt{nicerl2}. These filters result in a total filtered exposure time of $\sim$\,\,428\,\,ksec.

In order to create background-subtracted light curves, we used the \texttt{nibackgen3C50} tool, version 7 (\citealt{remillard2022}) to estimate the background. We used XSPEC version 12.12.1 (\citealt{arnaud1996}) for the spectral analysis. We grouped the spectra using the ``optimal binning'' scheme (\citealt{kaastra2016}), with a minimum of 25 counts per energy bin, using the ftools \texttt{ftgrouppha}. The detector response files were generated with the tasks \texttt{nicerarf} and \texttt{nicerrmf}, respectively.

Under the assumption that mHz QPOs are detected at low energies (see, e.g., \citealt{altamirano2008mQPOs}), we produced 1-s time resolution light curves in the 0.5--5.0 keV energy range for each NICER dataset with the tool \texttt{xselect}. We then proceeded to search for mHz QPOs by means of the Lomb-Scargle periodogram (LSP; \citealt{lomb1976}; \citealt{scargle1982}; \citealt{press1992}) in each individual gap-free data segment. In a few instances, after having detected the mHz oscillations, and to fill the gaps in the data, we re-ran the task \texttt{nicerl2} either relaxing the parameter \texttt{nicersaafilt} to ``no'' or the range of \texttt{overonly\_range} to 0--10.\footnote{\url{https://heasarc.gsfc.nasa.gov/lheasoft/ftools/headas/nicerl2.html}} 
Each time we detected a thermonuclear X-ray burst, we applied the LSP to look for variability before and after each burst.
We determined the mHz QPO frequency, $\nu_{\rm QPO}$, as the frequency at which the power peaks in its respective periodogram.
To estimate the significance level of the detections, we followed \citet{press1992} by assuming white noise and taking into account the number of trials, i.e., the number of frequencies searched, which varies from data-segment to data-segment. All the detections reported in this work have a significance $>$ 4$\sigma$ (i.e., a false-alarm probability $<$ $3.2 \times 10^{-5}$), and a data length of at least 700 s.

We created the hardness-intensity diagram (HID) to investigate the relationship between the spectral state of the source and the occurrence of mHz oscillations. Following \citet{Bult2018}, we produced 16 s binned light curves in the 3.8--6.8 and 2.0--3.8 keV energy bands. 
We defined the intensity as the count rate in the energy range 0.5--6.8 keV, and calculated the ``hard colour'' as the count rate ratio in the range 3.8--6.8 and 2.0--3.8 keV. In order to compute the hard colour and the intensity, we removed the type-I X-ray bursts. 

To study the energy dependence of the mHz QPOs, we folded each light curve in which we detected the oscillations in different energy bands using its measured frequency. We then fitted each folded light curve with a model consisting of a sinusoidal function plus a constant. We calculated the fractional rms amplitude of the oscillations within a data segment as a function of the energy through the formula $rms = A/[\sqrt{2} \times (C-B)]$, where $A$ is the sinusoidal amplitude of the oscillations, $C$ the constant persistent level, and $B$ the estimated X-ray background count rate.

%
%
%
%
%
%

\begin{figure*} 
\centering
\resizebox{2.07\columnwidth}{!}{\rotatebox{0}{\includegraphics[clip]{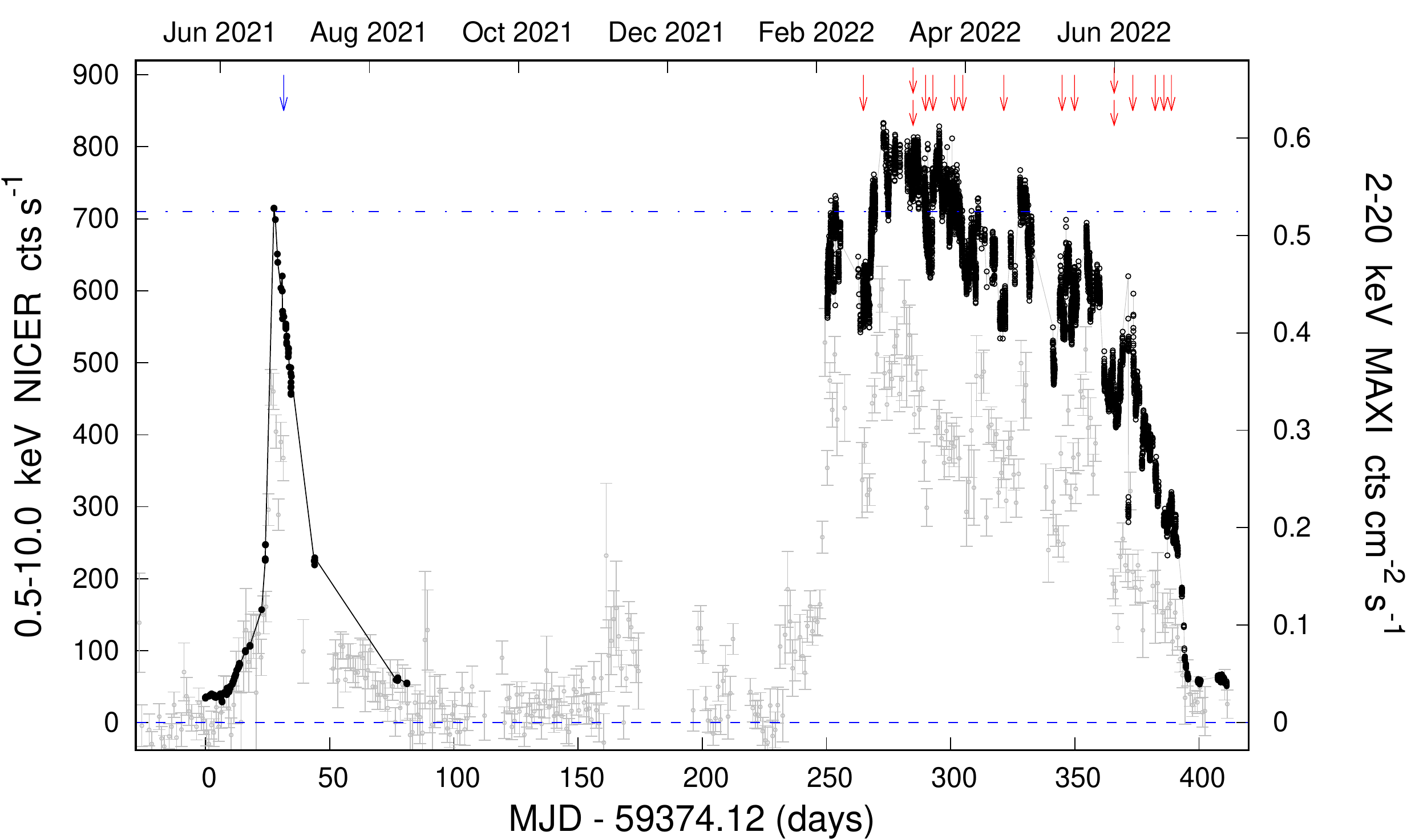}}}
\caption{Temporal evolution of the 2021 and 2022 outbursts of 4U~1730--22. Each black data point represents the count rate in the 0.5--10 keV band of NICER binned to 256 s for the case of the 2021 outburst, and to 32 s for the 2022 outburst. Each grey data point corresponds to the average MAXI daily count rate per unit area in the 2--20 keV range. Type-I X-ray bursts were removed for clarity and marked with a blue arrow for the only burst detected in 2021, and with red arrows for those detected in 2022. $t$ = 0 corresponds to the trigger of the NICER campaign. The blue dash-dotted horizontal line marks the peak count rate of the 2021 outburst.}
\label{fig:outbursts-together}
\end{figure*}

%
%
%
%
\section{Results}\label{sec:results}

\subsection{Outburst evolution}

In Fig.~\ref{fig:outbursts-together} we show the long-term light curves of both the 2021 and the last 2022 outburst of 4U~1730--22 with NICER (0.5--10 keV) and with MAXI (2--20 keV; \citealt{matsuoka2009}). During the 2021 outburst, NICER observed the source between 9 June (MJD 59374) and 29 August (MJD 59455). The source showed a relatively constant X-ray flux at $\sim$~37~cts~s$^{-1}$ over the first six days from the beginning of the campaign. Then the flux started increasing smoothly for several days. At around day 22, the intensity rose more quickly from $\sim$ 150 cts s$^{-1}$ to $\sim$ 280 cts s$^{-1}$ in two days. After a data gap of three days, the source flux decreased roughly monotonically from $\sim$ 710 cts s$^{-1}$ down to $\sim$ 470 cts s$^{-1}$ in a week. At approximately day 30, the first thermonuclear X-ray burst was observed from this source (see the blue arrow in Fig.~\ref{fig:outbursts-together} and \citealt{bult2021ATel}). Only five observations were performed after that during the next 37 days. The source flux was $\sim$ 60 cts s$^{-1}$, suggesting that the system was returning to quiescence.

Almost six months later, the MAXI nova alert system (\citealt{negoro2010}; nova ID 9623521891) reported a second outburst from 4U~1730--22 on 13 February 2022. NICER started observing the 2022 outburst of the source on 14 February 2022 (MJD 59624) until 18 August 2022 (MJD 59809; but see Sect.~\ref{sec:dataanalysis}). During the 2022 outburst (see Fig.~\ref{fig:outbursts-together}), the onset of the NICER campaign began when the source was already at an intensity of $\sim 600$ cts s$^{-1}$. Throughout the first 17 days of observations, the average X-ray flux of 4U~1730--22 decreased continuously. In this period, the first type-I X-ray burst was detected (see the red arrows in Fig.~\ref{fig:outbursts-together}). After that, the count rate increased from below $\sim 600$ cts s$^{-1}$ up to $\sim 800$ cts s$^{-1}$. The source remained at this X-ray level for approximately three weeks, although a noticeable drop in intensity at day $\sim$\,290 is observed. During this time interval, the system exhibited several X-ray bursts. A detailed analysis of the 17 X-ray bursts is presented in \citet{Bult2022}. From about day 293 onward, the flux decreased slowly but steadily albeit with multiple intensity increases until day 390. Afterwards, 4U~1730--22 declined to a low luminosity regime ($\lesssim$ 40 cts s$^{-1}$) in $\sim$ 5 days.

In Fig.~\ref{fig:HID}, we show the HID for both the 2021 and 2022 outbursts of 4U~1730--22. The source was sampled in two distinctive states. A high luminosity (with count rates $\gtrsim$~250~cts~s$^{-1}$) and soft (with hard colours between $\sim$\,0.24 and $\sim$\,0.32) state, and a low luminosity ($\lesssim$ 50 cts s$^{-1}$) but harder (with hard colours $\gtrsim$\,0.30) state. These two different states were identified by \citet{Bult2022} as a bright soft state and a faint hard state.

\subsubsection{One or two outbursts?}\label{sec:outbursts}

4U~1730--22 showed a fast decrease of the flux from its 2021 peak outburst in around 40 days, indicating that the source was returning to quiescence. NICER observed the system until 29 August 2021. After this date, the source was not observed with pointed observations of instruments such as XMM-Newton, Chandra, NuSTAR or Swift, at least until the beginning of the bright state in February 2022. The MAXI all-sky monitoring instrument observed 4U~1730--22 when it was in a very low luminosity phase (see filled grey points in Fig.~\ref{fig:outbursts-together}) between September 2021 and February 2022. These MAXI observations showed evidence of X-ray flux variability, particularly in November 2021 (around day 160 in Fig.~\ref{fig:outbursts-together}). Thus, it cannot be ruled out that this re-brightening was the actual onset of the second outburst (the one we called ``2022 outburst''). However, the large uncertainties in the data and the lack of supporting X-ray observations prevents us from confirming this result.
We are then unable to conclusively determine whether the source was ever in quiescence between September 2021 and February 2022 due to the lack of pointed observations. If 4U~1730--22 underwent only one outburst beginning in June 2021, this would indicate that this outburst was the longest ever recorded for this source.

\subsection{Detection of mHz QPOs}

We looked for the presence of mHz oscillations in the whole NICER data set. We did not find any significant cases of mHz QPOs in the 2021 outburst. Throughout the 2022 outburst, we detected 45 instances of mHz QPOs with significances greater than 4$\sigma$ and in data sets longer than 700~s in a total of 35 observations. The mHz QPO frequency was approximately constant within a data segment and between $\sim$ 4.5 and 8.1~mHz in the full data set. This is consistent with the frequency range of the mHz oscillations reported in other sources (e.g., \citealt{revnivtsev2001}). The data segments in which we observed the oscillations last between $\sim$ 0.7~ksec and up to $\sim$ 2.0 ksec. We did not find any evidence of systematic frequency drift within any of the individual data segments, probably due to the relatively short length of the data sets where we detected the mHz QPOs.  

\begin{figure}
\includegraphics[height=7.1cm,width=0.99\columnwidth]{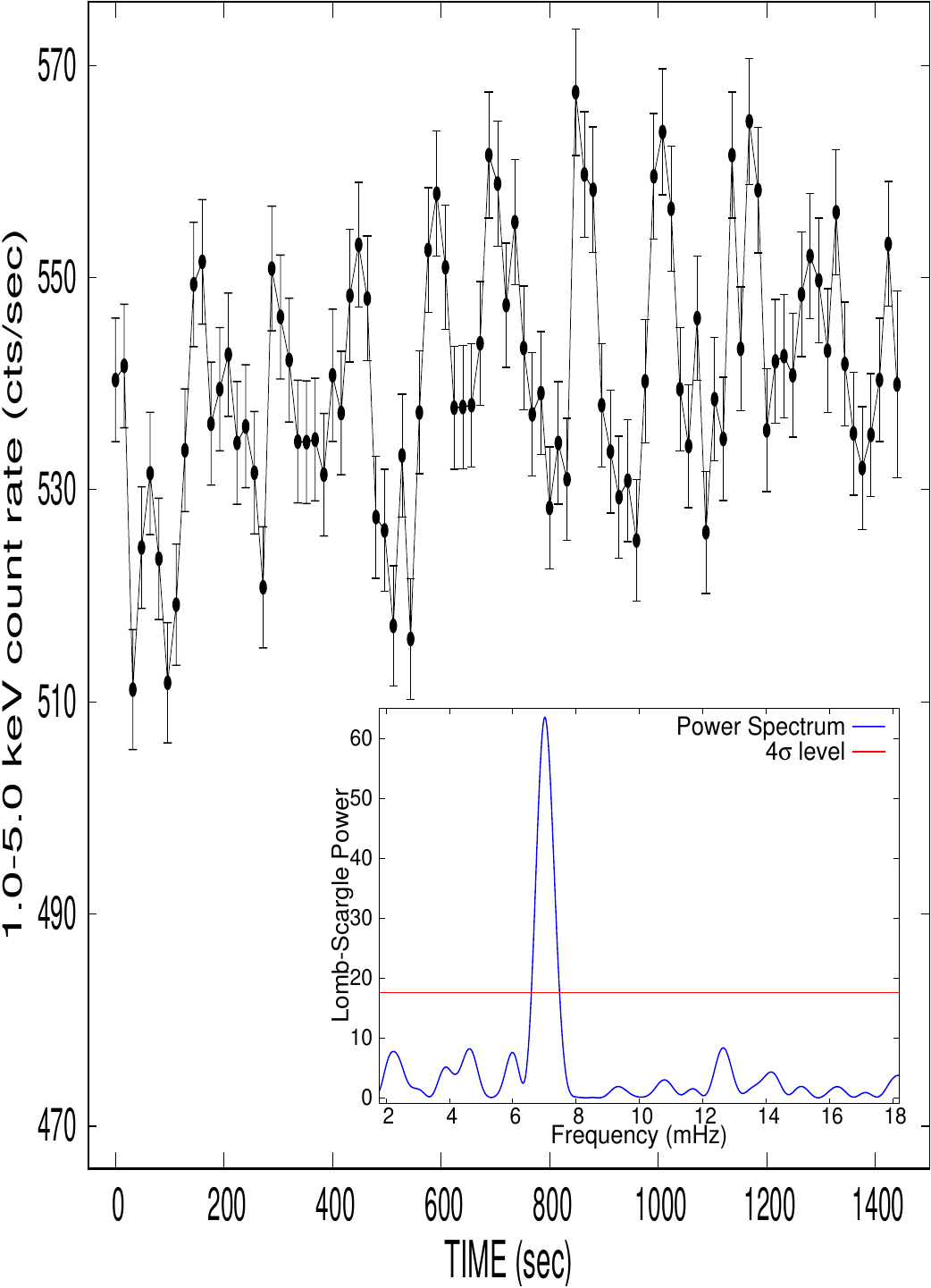}
\centering
\caption{NICER background-subtracted light curve (obsID 4639010144) in the 1.0--5.0 keV energy range of 4U~1730--22 rebinned to 16 s of an observation showing mHz QPOs. The corresponding Lomb–Scargle periodogram is shown in the inset. A prominent QPO is observed at a frequency of $\simeq$ 7 mHz.}
\label{fig:Fig-4639010144}
\end{figure}

\begin{figure}
\includegraphics[height=7.1cm,width=0.99\columnwidth]{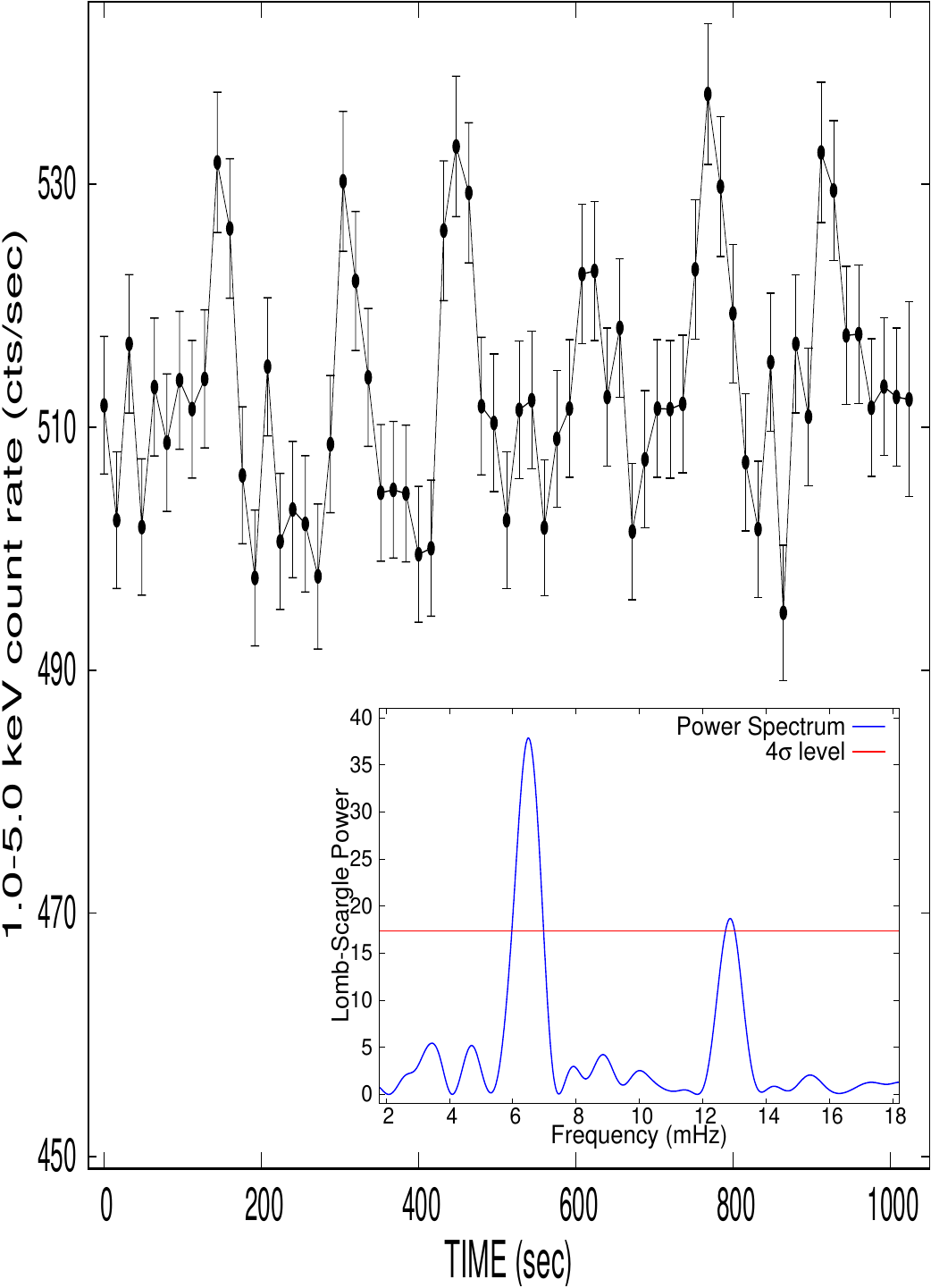}
\centering
\caption{NICER background-subtracted light curve (obsID 4202200143) in the 1.0--5.0 keV energy range of 4U~1730--22 rebinned to 16 s of an observation showing mHz QPOs. The corresponding Lomb–Scargle periodogram is shown in the inset. Note not only the significant QPO observed at a frequency of $\simeq$~6.4~mHz, but also the possible harmonic component at $\simeq$~12.8~mHz.}
\label{fig:Fig-4202200143}
\end{figure}

In Figs. \ref{fig:Fig-4639010144} \& \ref{fig:Fig-4202200143}, we show two representative examples of background-subtracted light curves (obsID 4639010144 and 4202200143) of segments in which we detected mHz QPOs. In both cases, we also show in the insets each corresponding LSP. In the two instances, a prominent QPO is observed at $\sim$\,7 mHz and $\sim$\,6.4 mHz, respectively. In the second example, a possible harmonic is also detected at $\sim$\,12.8 mHz. A similar result was reported by \citet{fei2021}, who found a harmonic component in a limited number of cases of data segments with mHz QPOs in 4U~1636--53. 

\begin{figure}
\includegraphics[height=7.1cm,width=0.99\columnwidth]{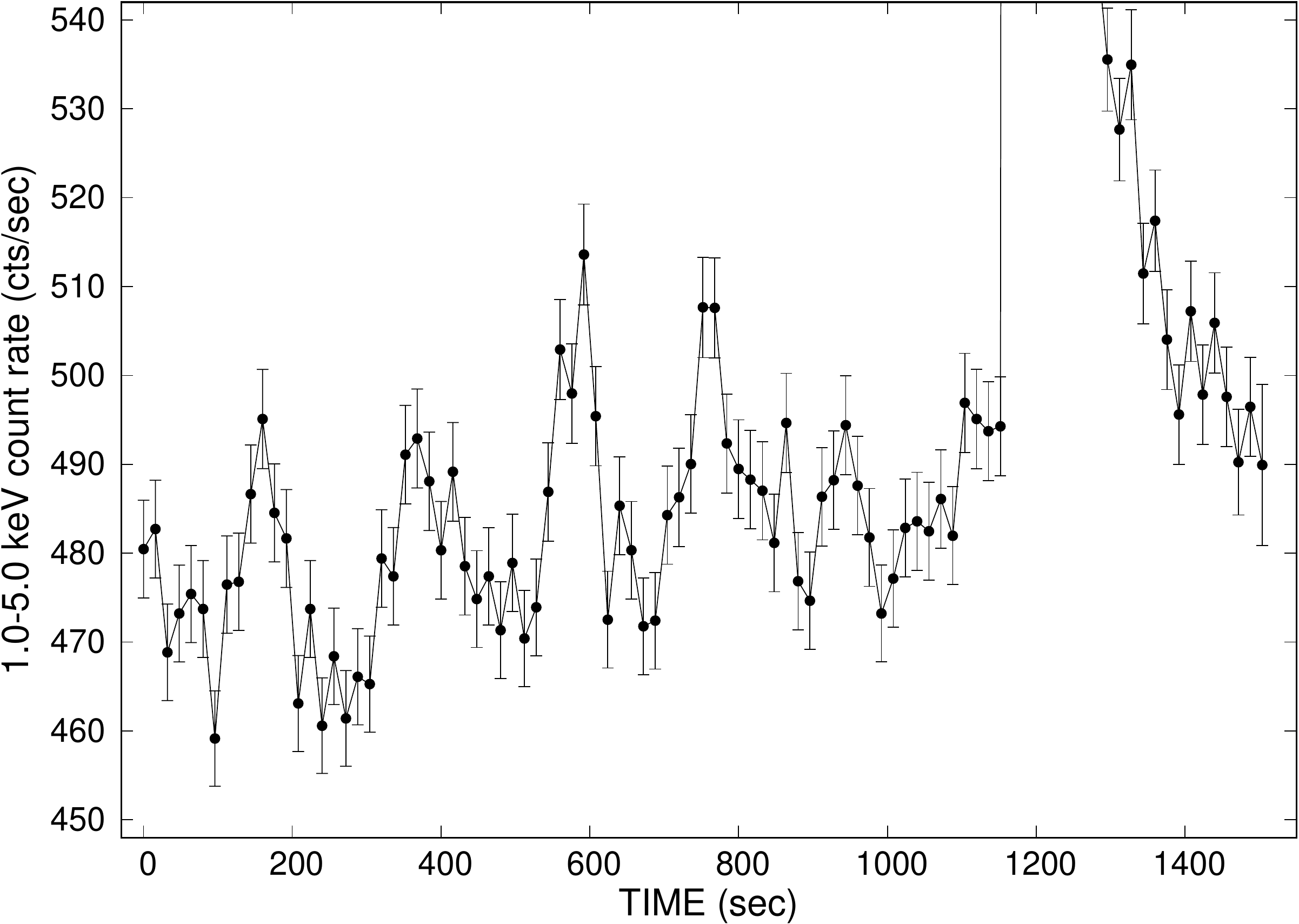}
\centering
\caption{Background corrected light curve 16\,\,s binned in the energy band 1.0--5.0\,\,keV of the NICER observation 4639010131 of 4U~1730--22. The plot shows mHz oscillations for around $\sim$\,1 ksec until the occurrence of a type-I X-ray burst.}
\label{fig:mQPOs-burst}
\end{figure}

We also found the mHz QPOs prior to the occurrence of a type-I X-ray burst in two cases. We show a representative light curve in Fig. \ref{fig:mQPOs-burst}. We did not find evidence of the oscillations after the bursts in the $\sim$ 200 sec of data available after the end of the bursts.

In Fig.~\ref{fig:rms-together} we plot two representative examples of the energy dependence of the fractional rms amplitude. We found an increasing trend of the rms amplitude with energy up to $\sim$ 3 keV in some cases (see left panel of Fig.~\ref{fig:rms-together}), whereas in others, the energy dependence was consistent with being constant (see right panel in Fig.~\ref{fig:rms-together}). The averaged fractional rms amplitude was $\sim$ 2--3\,\%. It is worth mentioning that we could only calculate upper limits (at 99.7\% confidence level) of the rms amplitude at energies larger than $\sim$ 3.5 keV.


\begin{figure*} 
\centering
\resizebox{1.6\columnwidth}{!}{\rotatebox{0}{\includegraphics[clip]{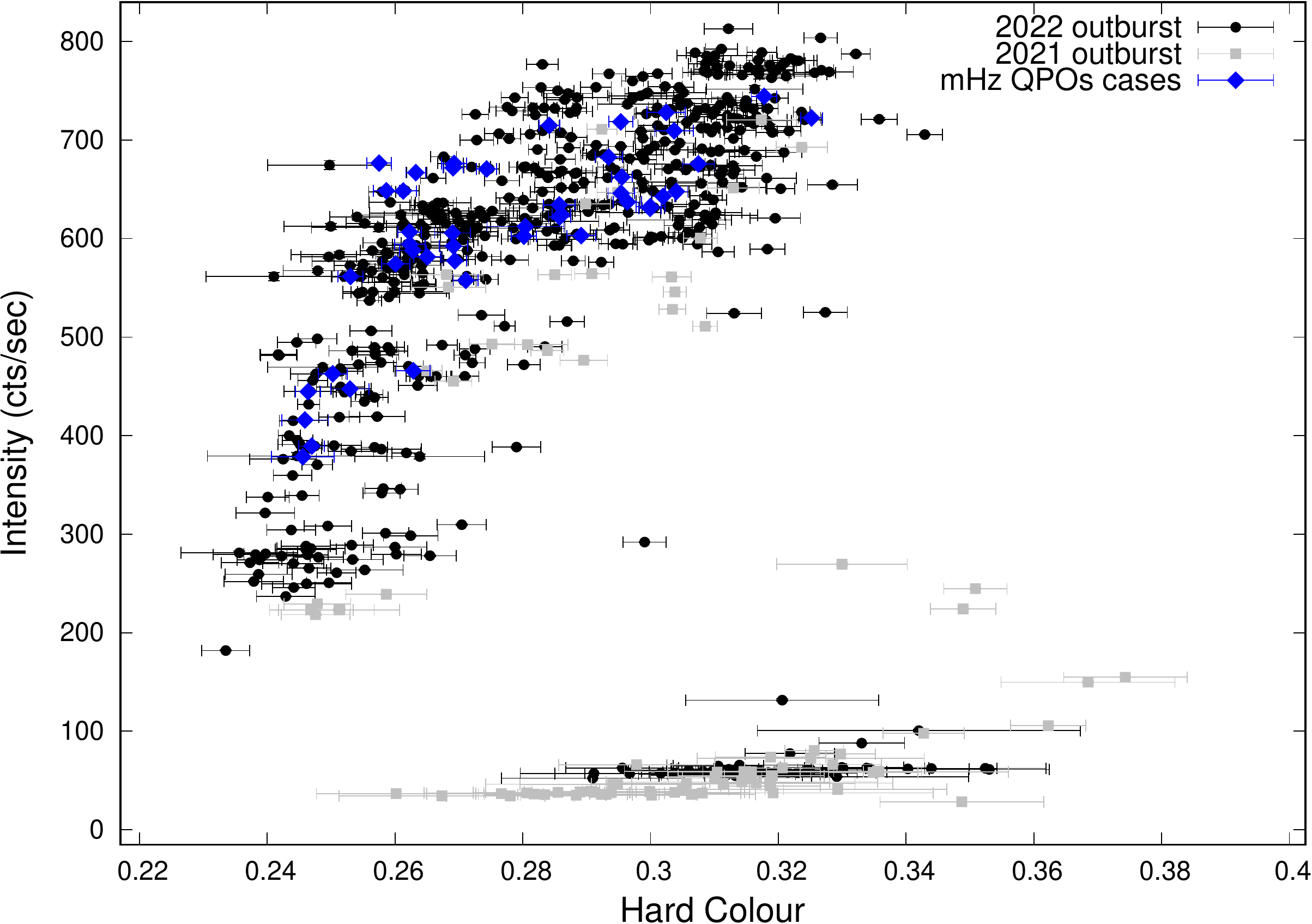}}}
\caption{Hardness-intensity diagram of 4U~1730--22 for all NICER data obtained during both the 2021 and 2022 outbursts. The hard colour is defined as the count rate ratio between the bands 3.8--6.8 and 2.0--3.8 keV, while the intensity is computed as the 0.5--6.8 keV count rate. Each filled grey square represents the hardness and the intensity of the source averaged per orbit from the 2021 outburst, whereas each solid black circle indicates those from the 2022 outburst. We also marked the mHz QPOs detections with blue diamonds.}
\label{fig:HID}
\end{figure*}


%
%
%

\section{Discussion}

\subsection{The origin of the mHz QPOs}

We report the discovery of mHz QPOs in the NS LMXB 4U~1730--22 during its 2022 outburst. The properties of the mHz QPOs that we observed in 4U~1730--22 are consistent with those found in other sources, namely, 4U~1636--53, 4U~1608--52, Aql~X--1, 4U~1323--619, GS~1826--238, EXO~0748--676, and 1RXS~J180408.9--342058.

We detected the mHz oscillations with frequencies between $\sim$\,4.5 and $\sim$\,8.1 mHz. These frequencies are within the range of QPO frequencies reported in the aforementioned sources ($\sim$\,4--14 mHz; \citealt{revnivtsev2001}; \citealt{strohmayer2011}; \citealt{strohmayer2018}; \citealt{Mancuso2019}; \citealt{tse2021}; \citealt{mancuso2021}). More importantly, we found two cases in which the mHz oscillations were followed by a thermonuclear (type-I) X-ray burst and disappeared afterwards (see Fig. \ref{fig:mQPOs-burst}). This is one of the main characteristics that separates mHz QPOs from other types of QPOs observed in NS systems. These mHz QPOs are interpreted as the marginally stable nuclear burning of He on the NS surface (\citealt{heger2007}).

We explored the relationship between the source spectral state and the occurrence of the mHz QPOs through the HID. We detected the mHz QPOs when the source was in a bright and soft spectral state (\citealt{Bult2022}) during the NICER observations (see Fig.~\ref{fig:HID}). The soft state identification is supported by the fact that the source spectrum was well described by an absorbed two-thermal emission component (a disc blackbody and a blackbody model) with temperatures $kT_{\rm in} \simeq$ 0.8 keV and $kT_{\rm bb} \simeq$ 1.6 keV (see below).
A full spectral analysis of 4U~1730--22 is beyond the scope of this paper and has already been performed by \citet{chen2022}.
Our model is consistent with those observed during the soft state in other sources, in which thermal components are generally predominant (see, e.g., \citealt{lin2007} and references therein; \citealt{tarana2011}). Therefore, we infer that we found the mHz QPOs when 4U~1730--22 was in a soft spectral state. This is in agreement with the spectral colours of the mHz QPO detections in other systems\footnote{In some instances, the oscillations were detected at the transition between the soft and intermediate states (\citealt{altamirano2008mQPOs}; \citealt{Mancuso2019}; \citealt{mancuso2021}). In the case of 1RXS~J180408.9--342058, the source was in a hard spectral state (\citealt{tse2021}).} (\citealt{revnivtsev2001}; \citealt{altamirano2008mQPOs}; \citealt{strohmayer2018}; \citealt{Mancuso2019}; \citealt{lyu2019}; \citealt{mancuso2021}).

\begin{figure*} 
\centering
\resizebox{2.05\columnwidth}{!}{\rotatebox{0}{\includegraphics[clip]{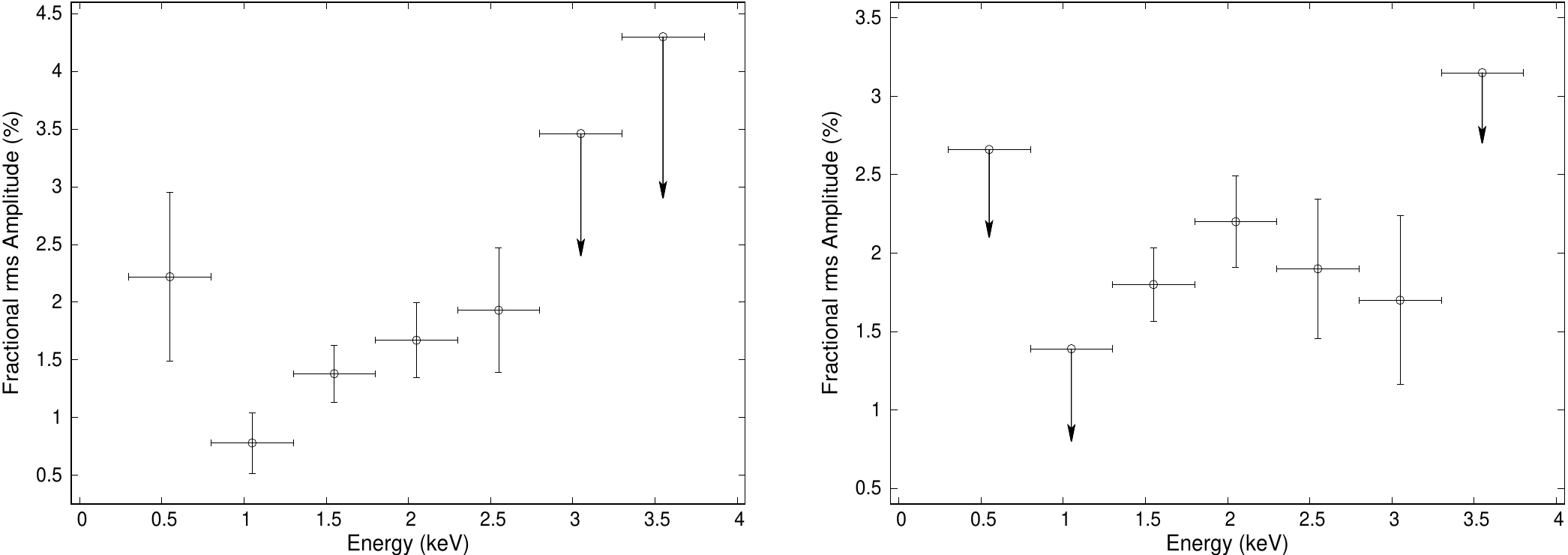}}}
\caption{Representative examples of the fractional rms amplitude versus energy of the mHz oscillations observed with NICER in 4U~1730--22. The points represent the middle energy (band) whereas the horizontal error bar the whole energy band used in each case. The left panel shows an increasing tendency of the rms towards higher energies up to $\sim$\,3 keV. Instead, the right panel shows an apparent flattening of the rms amplitude with increasing energy below $\sim$\,3 keV.}
\label{fig:rms-together}
\end{figure*}

We studied the energy dependence of the fractional rms amplitude of the mHz pulses in 4U~1730--22. Fig. \ref{fig:rms-together} (left panel) shows a slight increase of the fractional rms amplitude with energy, from 1~keV up to $\sim$ 3~keV. A similar behaviour was previously reported by \citet{strohmayer2018} in observations of GS~1826--238 with NICER and by \citet{lyu2020} in observations of 4U~1636--53 with NICER and XMM-Newton. In both works, a clear increasing tendency of the rms amplitude of the mHz QPOs was found below $\sim$ 3~keV. Our results support the speculation of \citet{lyu2020} that the increasing trend up to $\sim$ 3 keV is an intrinsic characteristic of the mHz QPOs.\footnote{Note that in 1RXS~J180408.9--342058, the opposite trend was reported.} For the rest of the sources (4U~1608--52, Aql~X--1, 4U~1323--619, and EXO~0748--676), such feature could not be confirmed given that the oscillations were mainly observed with RXTE, which operated at energies above 2~keV. We also found instances in which the trend is less clear. In some cases, the rms spectra were compatible with being constant with energy below $\sim$ 3.5 keV. In others, a modest increase below $\sim$ 2 keV and a decrease at higher energies could not be discarded either (see right panel in Fig.~\ref{fig:rms-together}). The large uncertainties prevented us from reaching a firm conclusion.

For energies above $\sim$ 2.5 keV and up to $\sim$ 5 keV, a systematic decrease of the rms amplitude has been observed in 4U~1636--53 and 4U~1608--52 (\citealt{revnivtsev2001}; \citealt{altamirano2008mQPOs}; \citealt{lyu2020}). For the case of GS~1826--238, \citet{strohmayer2018} reported a flattening of the rms amplitude above $\sim$ 3 keV. In 4U~1730--22, we could only calculate upper limits (at 99.7\% confidence level) of the rms amplitude at energies $\gtrsim$ 3.5 keV, with values of $\sim$ 3--4\,\%. This might be due, on the one hand, to the reduction of the effective area of the NICER detectors at energies above $\sim$ 3 keV and on the other hand, to the fall in intensity of the system at high energies. 
Therefore, given the large uncertainties, we are unable to test whether our results are different from those reported in the aforementioned sources. 

To estimate the luminosity at which the oscillations occurred, we identified the data segments in which the QPOs were present and had the highest and lowest flux in the 0.5--10 keV energy band. Then, we selected the two data segments closest in both flux and time to the highest/lowest flux segments that did not contain QPOs. We avoided the data segments with mHz oscillations since models predict a modulation of the temperature of the NS surface along the mHz cycle (\citealt{heger2007}), which might significantly affect the spectral fittings (but see below). We finally extracted the respective source and background spectrum via the \texttt{nibackgen3C50} tool, and created the response and ancillary files (see Sect. \ref{sec:dataanalysis}). The spectra are reasonably well described by two thermal components, a black-body (\texttt{bbodyrad} in XSPEC) plus a multi-temperature black-body disc (\texttt{diskbb} in XSPEC; \citealt{mitsuda1984}; \citealt{makishima1986}) models. We used \texttt{phabs} to take into account the interstellar absorption along the line of sight, together with ``wilm'' abundances (\citealt{wilms2000}) and ``vern'' cross-sections (\citealt{verner1996}). We multiplied our model by the convolution model \texttt{cflux} to calculate the extrapolated unabsorbed flux in the 2--20 keV energy range. To compute the luminosities, we used the distance estimated by \citet{Bult2022}, i.e., $d = 6.9$ kpc. We found that the X-ray luminosity was constrained to $L_{\rm X} \simeq (0.5 - 1.2) \times 10^{37}$~erg~s$^{-1}$ (these estimates do not take the uncertainties in the distance). Assuming an Eddington luminosity of $L_{\rm Edd}$ = 3.8 $\times$ 10$^{38}$~erg~s$^{-1}$ (\citealt{kuulkers2003}) for a canonical NS of 1.4 $M_{\odot}$, our results suggest that the mHz QPOs occurred when 4U~1730--22 had a luminosity of 1--4\%\,$L_{\rm Edd}$.
These values of the luminosity at which the mHz oscillations are seen are compatible with previous works on other sources (\citealt{revnivtsev2001}; \citealt{altamirano2008mQPOs}; \citealt{strohmayer2018}; \citealt{Mancuso2019}; \citealt{tse2021}). The fact that we detected the mHz QPOs in a relatively narrow range of luminosities would explain why we did not find any significant case of QPOs during the previous outburst. In the 2021 outburst, the source spent a small fraction of the observed time at high enough luminosities at which the QPOs might have been detected (see Fig.~\ref{fig:HID}). 

Another alternative explanation for the non-detection of mHz QPOs during the 2021 outburst would be the following. \citet{cavecchi2020} studied how the burst rate changes with the accretion rate in five LMXBs. They observed a consistent pattern across all five sources: the burst rate initially increased to reach a peak (the ``first branch''), and then declined as the accretion rate increased (the ``second branch''). The critical mass accretion rate at which the burst rate started decreasing was anticorrelated with the NS spin frequency. The authors attributed the decrease in the burst rate to the stabilisation of the burning on a growing portion of the NS surface. They suggested that the stabilisation would begin probably first around the equator, and then move towards other latitudes. In addition, \citet{cavecchi2020} noted that, for 4U 1636--53, the mHz QPOs appeared right before the onset of the stabilisation, and were present in the second branch. If we assume that 4U~1730--22 behaves similarly to the sources studied by \citet{cavecchi2020}, we can explain the absence of mHz QPOs during the 2021 outburst as the source being in its first branch, with local mass accretion rates below the critical value for the stabilisation. Throughout the 2022 outburst, either different local conditions for the mass accretion rate or a different fuel composition along the latitude (due to the effect of rotationally induced mixing) may have led to an early stabilisation of a fraction of the NS and hence, the appearance of the oscillations.

We attempted to carry out phase-resolved spectroscopy of the mHz QPOs to further investigate the possible connection between the oscillations and the predicted modulation of the temperature of the NS surface (\citealt{heger2007}). However, the relatively short length of the mHz QPOs data segments sample prevented us from obtaining any meaningful result due to an insufficient signal-to-noise ratio. Even if adding and averaging all the mHz QPOs cases could have been a way to improve the signal-to-noise ratio, the fact that the oscillations were observed at different persistent emission levels, implying changes in the component models and/or in their parameter values, impeded performing the analysis.

In conclusion, based on all the similarities described above between the observational properties of the mHz oscillations seen in 4U~1730--22 and in the previously reported sources, we conclude that the origin of the mHz QPOs we observed in this source is very likely the same as in the other systems, i.e., due to MSNB of He in the NS surface (\citealt{heger2007}). This makes 4U~1730--22 the eighth source which shows QPOs associated with the MSNB phenomenology. 

Up to date, all the previous systems which exhibited mHz QPOs are either persistent sources or underwent multiple outbursts in the last decades (see Sect.~\ref{sec:intro}). This is the first time we observe mHz QPOs in a system which had been in quiescence for almost 50 years (see Sect.~\ref{sec:outbursts}). It then opens a question if any NS LMXB, independently of its accreting history, might reach the necessary conditions on the NS surface for MSNB. For example, the heat flux from the NS crust could affect the frequency of the oscillations making it decrease (\citealt{keek2009}), but might not be a factor for the onset of the mHz QPOs. In any case, more theoretical work will be required to better understand how the key parameters (such as the NS surface temperature or the density) govern the occurrence of the mHz QPOs.


%
%
%

\vspace{-0.25cm}
\subsection{Summary}

We report the discovery of mHz quasi-periodic oscillations during the last 2022 outburst in the NS LMXB 4U~1730--22. Our findings are consistent with those found in other sources, and therefore strongly suggest that the mechanism responsible for the mHz QPOs is shared among the different systems. This makes 4U~1730--22 the eighth source showing mHz QPOs associated with MSNB, together with 4U~1636--53, 4U~1608--52, Aql~X--1, 4U~1323--619, GS~1826-238, EXO~0748--676, and 1RXS~J180408.9--342058.

\vspace{-0.35cm}
\section*{Acknowledgments}

GCM acknowledges support from the Royal Society International Exchanges ``the first step for High-Energy Astrophysics relations between Argentina and the UK''. 
DA acknowledges support from the Royal Society. 
GCM was partially supported by PIP 0113 (CONICET). This work received financial support from PICT-2017-2865 (ANPCyT). 
PB acknowledges support from NASA through the Astrophysics Data Analysis
Program (80NSSC20K0288) and the CRESST II cooperative agreement (80GSFC21M0002).
SG acknowledges the support of the CNES.
TG has been supported in part by the Turkish Republic, Presidency of Strategy and Budget project, 2016K121370.

This research has made use of data and/or software provided by the High Energy Astrophysics Science Archive Research Center (HEASARC), which is a service of the Astrophysics Science Division at NASA/GSFC and the High Energy Astrophysics Division of the Smithsonian Astrophysical Observatory. This research has made use of NASA's Astrophysics Data System. This research has made use of the MAXI data provided by RIKEN, JAXA and the MAXI team.

\vspace{-0.15cm}
\section*{Data availability}

The data underlying this article are publicly available in the High Energy Astrophysics Science Archive Research Center (HEASARC) at \url{https://heasarc.gsfc.nasa.gov/db-perl/W3Browse/w3browse.pl}.

\vspace{-0.15cm}
\bibliographystyle{mnras}
\bibliography{biblio2.bib}

\vspace{5mm}

\label{lastpage}
\end{document}